\documentclass[useAMS,usenatbib]{mn2e}
\usepackage[dvips]{epsfig}
\usepackage[]{natbib}
\usepackage[]{fixltx2e}
\usepackage[]{amsmath}
\usepackage[]{amssymb}
\title[Feeding and Feedback in NGC\,3081]{Feeding and Feedback in NGC\,3081}
\author[A. Schnorr-M\"uller et al.]
  {Allan Schnorr-M\"uller,$^{1,}$$^2$  Thaisa Storchi-Bergmann,$^3$ Andrew Robinson,$^4$ Davide Lena,$^{5,6}$ 
  \newauthor Neil M. Nagar$^7$\\
  $^1$Max-Planck-Institut f\"ur Extraterrestrische Physik, Giessenbachstr. 1, D-85741 Garching, Germany\\
  $^2$CAPES Foundation, Ministry of Education of Brazil, 70040-020, Bras\'ilia, Brazil\\
  $^3$Instituto de F\'isica, Universidade Federal do Rio Grande do Sul, 91501-970, Porto Alegre, RS, Brazil\\
  $^4$Physics Department, Rochester Institute of Technology, Rochester, New York 14623, USA\\
  $^5$SRON Netherlands Institute for Space Research, Sorbonnelaan 2, NL-3584 CA Utrecht, the Netherlands\\
  $^6$Department of Astrophysics/IMAPP, Radboud University, Nijmegen, PO Box 9010, NL-6500 GL Nijmegen, the Netherlands\\
  $^7$Astronomy Department, Universidad de Concepci\'on, Casilla 160-C, Concepci\'on, Chile\\}

\pagerange{\pageref{firstpage}--\pageref{lastpage}} \pubyear{2016}

\begin{document}
\label{firstpage}

\maketitle

\begin{abstract}

We present two-dimensional gaseous kinematics of the inner 1.2\,$\times$\,1.8\,kpc$^2$ of the Seyfert\,2 galaxy NGC\,3081, from optical spectra (5600--7000\,\r{A}) obtained with the GMOS integral field spectrograph on the Gemini North telescope at a spatial resolution of $\approx$\,100\,pc. We have identified  two-components in the line emitting gas. A narrower component (FWHM $\approx$\,60-100\,km\,s$^{-1}$), which appears to be gas in the galaxy disk, and which shows a distorted rotation pattern, is observed over the whole field of view. A broader component (FWHM $\approx$150-250\,km\,s$^{-1}$) is present in the inner $\approx$\,2\arcsec (200\,pc) and shows blueshifts and redshifts in the near and far sides of the galaxy, respectively, consistent with a bipolar outflow. Assuming this to be the case, we estimate that the mass outflow rate in ionized gas ($\dot{M}_{out}$) is between 1.9\,$\times\,10^{-3}$M$_{\odot}$\,yr$^{-1}$ and 6.9\,$\times\,10^{-3}$M$_{\odot}$\,yr$^{-1}$. The subtraction of a rotation model from the narrower component velocity field reveals a pattern of excess blueshifts of $\approx$\,50\,km\,s$^{-1}$ in the far side of the galaxy and similar excess redshifts in the near side, which are cospatial with a previously known nuclear bar. We interpret these residuals as due to gas following non-circular orbits in the barred potential. Under the assumption that these motions may lead to gas inflows, we estimate an upper limit for the mass inflow rate in ionized gas of $\phi$\,$\approx$\,1.3\,$\times\,10^{-2}$M$_{\odot}$\,yr$^{-1}$.

\end{abstract}

 
\section{Introduction}

It is widely accepted that the radiation emitted by an active galactic nucleus (AGN) is a result of accretion on to the central supermassive black hole (hereafter SMBH). However, understanding how mass is transferred from kiloparsec scales down to nuclear scales has been a long-standing problem in the study of nuclear activity in galaxies.

Theoretical studies and simulations have shown that non-axisymmetric potentials efficiently promote gas inflow towards the inner regions \citep{shlosman90,emsellem03,knapen05,emsellem06}. Simulations by \citet{maciejewski04a,maciejewski04b} demonstrated that, if a central SMBH is present, spiral shocks can extend all the way to the SMBH vicinity and generate gas inflow consistent with the observed accretion rates. Simulations presented in \citet{hopkins10} showed that on scales of $\lesssim$\,500\,pc in gas rich systems a system of nested gravitational instabilities with a range of morphologies such as nuclear spiral arms, rings, barred rings, clumpy disks and streams generate gas inflows. Nuclear bars have also been identified as a mechanism capable of driving gas inwards to scales of tens of parsecs \citep{shlosman89,englmaier04}.

A strong correlation between the presence of nuclear dust structures (spirals, filaments and disks) and nuclear activity in early type galaxies was reported by \citet{lopes07}. Specifically, they found that there is a marked difference in the dust and gas content within the inner $\approx$\,1\,kpc of early-type active and non-active galaxies: while the former always have dusty structures, in the form of spiral and filaments on scales of hundreds of parsecs, only 25\% of the non-active ones have such structures. This indicates that a reservoir of gas and dust is a necessary condition for the nuclear activity and it also suggests that the dusty structures are tracers of feeding channels to the AGN. This is confirmed in the study of \citet{martini13} who found that in the AGN hosts the mass of dust ranges between 10$^{5}$ and 10$^{6}$M$_{\odot}$, implying 10$^{7}$--10$^{8}$M$_{\odot}$ of gas in the inner $\approx$\,1\,kpc.

In order to test the hypothesis that nuclear spirals channel gas inwards to feed the SMBH, our group has been mapping gas flows in the inner kiloparsec of nearby AGNs using optical and near-infrared integral field spectroscopic observations of the inner kiloparsec of nearby AGN. So far, we have observed gas inflows along nuclear spirals in NGC\,1097 \citep{fathi06}, NGC\,6951 \citep{thaisa07}, NGC\,4051 \citep{rogemar08}, M\,79 \citep{rogemar13}, NGC\,2110 (\citealt{allan14a}; \citealt{diniz14}) and NGC\,7213 \citep{allan14b}. We also observed gas inflows in the galaxy M\,81 \citep{allan11}, where the inflow was mostly traced by dust lanes. Hints for the presence of inflows along dust lanes were also found in NGC\,1386 \citep{lena15}. Gas inflows along nuclear spirals have also been observed by other groups. Near-infrared integral field spectroscopic observations revealed inflows along nuclear spiral arms in NGC\,1097 \citep{davies09} and in NGC\,7743 \citep{davies14}. Recent ALMA (Atacama Large Millimeter Array) observations revealed streaming motions along nuclear spirals in NGC\,1433 \citep{combes13} and NGC\,1566 \citep{combes14}.

Inflows driven by large-scale stellar bars have been observed in many objects, for example, NGC\,4151 \citep{mundell99}, NGC\,4569 \citep{boone07}, NGC\,6951 \citep{laan11} and NGC\,3227 \citep{davies14}. It has been found that, among local early-type barred spirals, $\approx$\,30\% host an inner secondary bar \citep{erwin02,laine02,erwin04}, suggesting that such systems are common.  Gas inflows driven by nuclear bars, however, have only been observed in a few objects so far \citep{caceres13}. The Nuclei of Galaxies (NUGA) team \citep {burillo03} found that in a sample of 25 local low luminosity AGNs, gas is being driven into the central 100\,pc in 1/3 of them \citep{haan09,combes12}.

In this work, we present an analysis of the gas kinematics and excitation in the inner kiloparsec of the active galaxy NGC\,3081 as derived from integral field spectroscopic observations carried out with the Gemini North telescope. NGC\,3081 is a S0/a galaxy hosting a Seyfert 2 nucleus. At a distance of 37.7\,Mpc (from NED\footnote{NASA/IPAC extragalactic database}) the resulting scale is 179\,pc/\arcsec (cosmology corrected assuming H$_{0}$\,=\,73\,km\,s$^{-1}$\,Mpc$^{-1}$, $\Omega_{\mathrm{matter}}$\,=\,0.27 and $\Omega_{\mathrm{vacuum}}$\,=\,0.73). 

NGC\,3081 has a weak large scale bar and a nuclear bar \citep{buta90}, and four resonance rings: two outer rings, an inner ring and a nuclear ring \citep{buta98}. Inside the nuclear ring, two spiral arms are observed. A comparison between the continuum adjacent to the [O\,III] and to the [N\,II]\,+\,H$\alpha$ lines showed that the nuclear ring and nuclear spiral arms are sites of recent or ongoing star formation \citep{ferruit00}. Radio observations revealed a compact radio source oriented roughly along the north-south direction \citep{nagar99}. Slitless spectroscopy by \citet{ruiz05} revealed the presence of an outflow in the inner 2\arcsec\ of the galaxy. It is also worth noting that broad emission lines have been detected in polarized light in this object -- see \citet{moran00}. However, broad near-infrared emission lines were not detected by \citet{goodrich94} or \citet{reunanen03}.

The present paper is organized as follows. In Section \ref{Observations}, we describe the observations and data reduction. In Section \ref{Results}, we present the procedures used for the analysis of the data and the subsequent results. In section \ref{Discussion}, we discuss our results and present estimates of the mass inflow rate and mass outflow rate and in Section \ref{Conclusion}, we present our conclusions.

\section {Observations and Data reduction}\label{Observations}

\begin{figure*}
\includegraphics[scale=0.8]{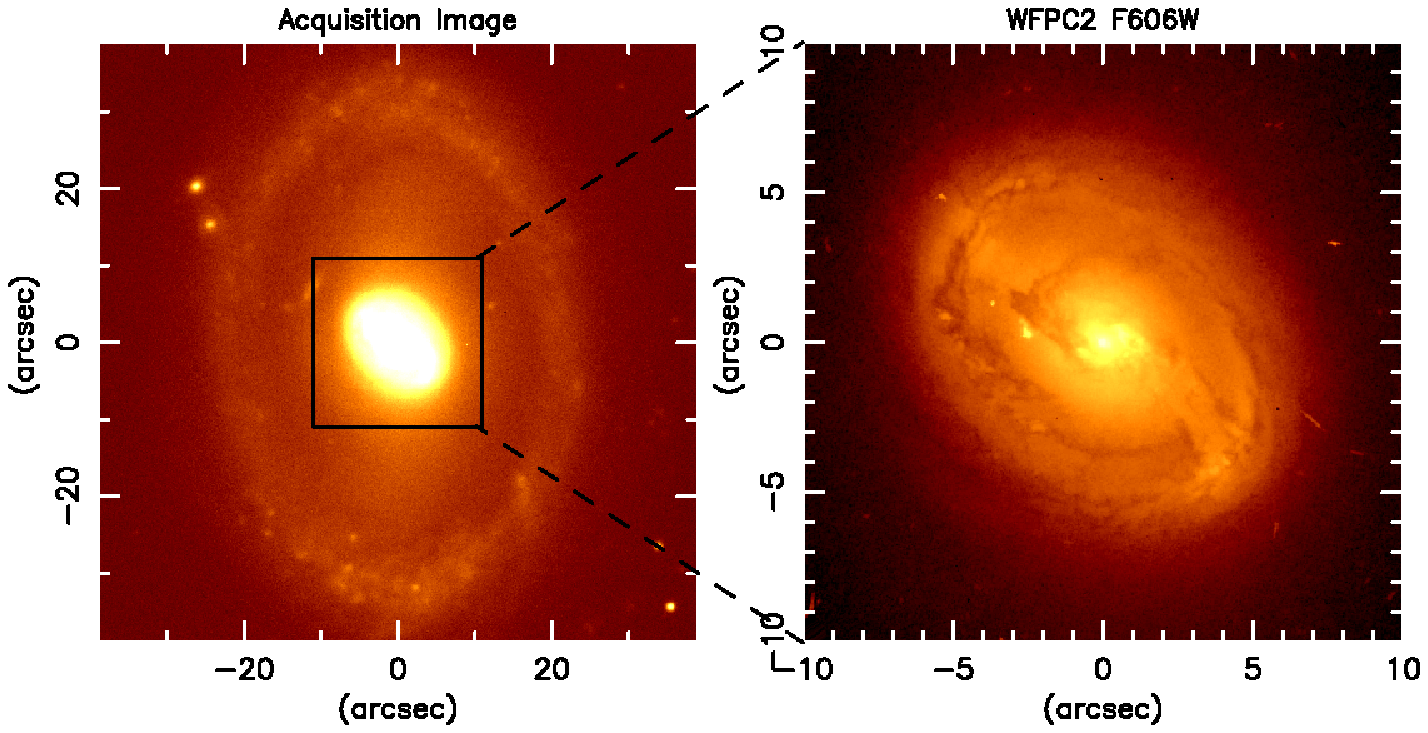}
\includegraphics[scale=0.8]{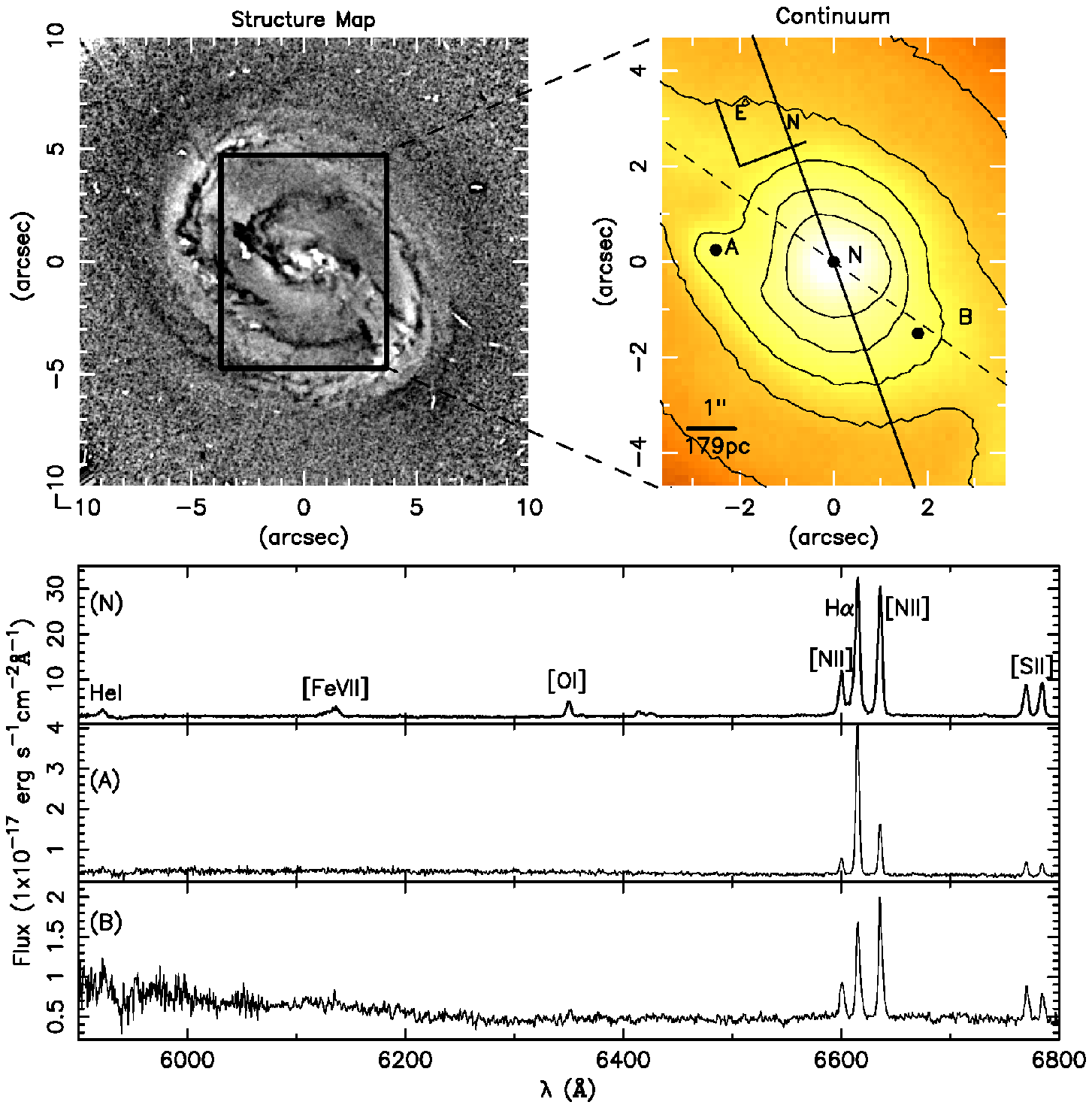}
\caption[Large scale image of NGC\,3081]{Top left: acquisition image. Top right: HST WFPC2 F606W image of the inner 20\arcsec\,$\times$\,20\arcsec. Middle left: structure map obtained from the HST image; the rectangle delimits the FOV of the GMOS-IFU observations. Middle right: continuum image extracted from the IFU spectra. The continuous black line indicates the position of the kinematic major axis (PA\,=\,90\ensuremath{^\circ}), as determined from the stellar velocity field in the inner 8\arcsec\ by \citet{stoklasova09}. The dashed black line indicates the position of the nuclear bar (PA\,=\,122\ensuremath{^\circ}). Bottom: spectra corresponding to the regions marked as N, A and B in the IFU image.}
\label{fig1}
\end{figure*}

The observations were obtained with the integral field unit of the Gemini Multi Object Spectrograph (GMOS-IFU) at the Gemini North telescope on the night of February 14, 2011 (Gemini project GN-2011A-Q-85). The observations consisted of two adjacent IFU fields (covering 7\,$\times$\,5\,arcsec$^{2}$ each) resulting in a total angular coverage of 7\,$\times$\,10\,arcsec$^{2}$ around the nucleus with the largest side of the field oriented along position angle (PA) 70\ensuremath{^\circ}. Four exposures of 615 seconds were obtained for each field, slightly shifted spatially and dithered spectrally in order to correct for detector defects and cosmic rays after combination of the frames. The seeing during the observation was 0\farcs6, as measured from the full width at half-maximum (FWHM) of a spatial profile of the calibration standard star. This corresponds to a spatial resolution at the galaxy of $\approx$\,100\,pc.

The selected wavelength range was 5600-7000\,\r{A}, in order to cover the H$\alpha$+[N\,II]\,$\lambda\lambda$6548,6583 and [S\,II]\,$\lambda\lambda$6716,6731 emission lines. The observations were obtained with the grating GMOS R400-G5305 (set to central wavelength of either $\lambda$6500\,\r{A} or $\lambda$6550\,\r{A}) at a spectral resolution of R\,$\approx$\,2000. The wavelength calibration is accurate to the order of 8\,km\,s$^{-1}$.

The data reduction was performed using specific tasks developed for GMOS data in the \textsc{gemini.gmos} package as well as generic tasks in \textsc{iraf}\footnote{\textit{IRAF} is distributed by the National Optical Astronomy Observatories, which are operated by the Association of Universities for Research in Astronomy, Inc., under cooperative agreement with the National Science Foundation.}. The reduction process comprised bias subtraction, flat-fielding, trimming, wavelength calibration, sky subtraction, relative flux calibration, building of the data cubes at a sampling of 0\farcs1$\,\times\,$0\farcs1, and finally the alignment and combination of eight data cubes.

\section{Results}\label{Results}

\begin{figure}
\includegraphics[scale=1.19]{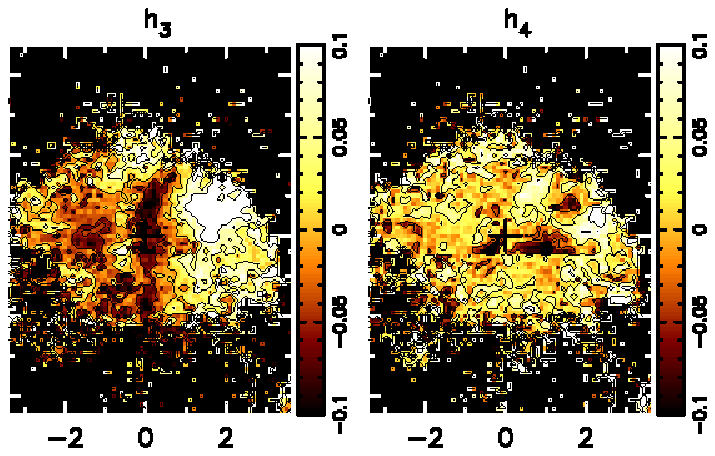}
\caption{Map of the Gauss-Hermite moments $h3$ and $h4$.}
\label{fig2}
\end{figure}

\begin{figure*}
\includegraphics[scale=0.8]{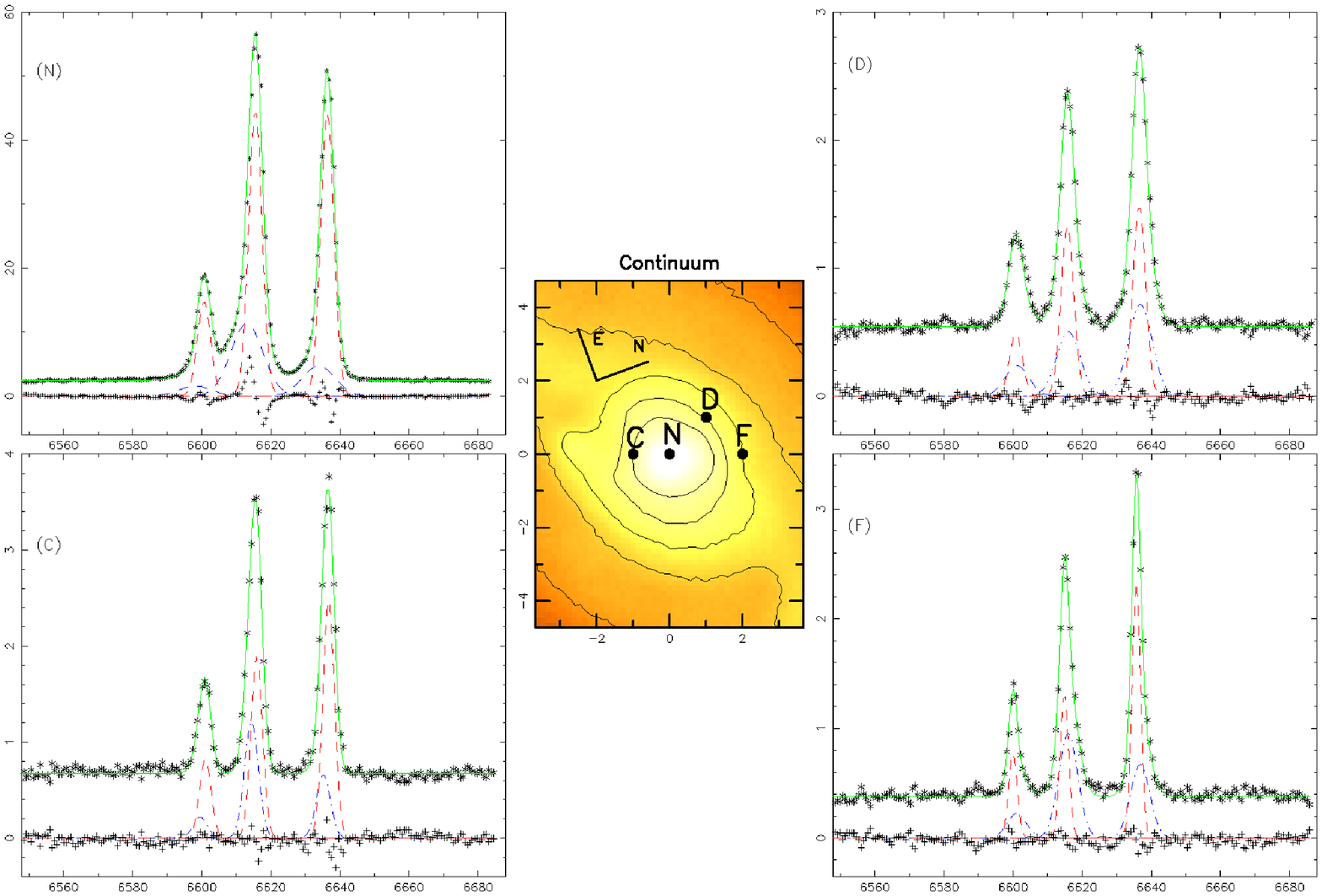}
\caption{Example of the two Gaussians fit for four different spectra.}
\label{fig3}
\end{figure*}

In Fig.\,\ref{fig1}, we present in the upper left panel the acquisition image of NGC\,3081 and in the upper right panel an image of the inner 22\arcsec\,$\times\,$22\arcsec\ of the galaxy obtained with the WFPC2 (Wide Field Planetary Camera 2) through the filter F606W aboard the Hubble Space Telescope (HST). In the middle left panel, we present a structure map of the HST image (see \citealt{lopes07}). Four features are visible in the structure map: a nuclear ring, a nuclear bar and two nuclear spirals arms (delineated by dark lanes) emerging from the nuclear bar. The rectangle delimits the field-of-view (hereafter FOV) covered by the IFU observations. In the middle right panel, we present a continuum image from the IFU spectra. The nuclear bar is visible along PA\,=\,122\ensuremath{^\circ} (dashed line, \citealt{erwin04}). In the lower panel, we present three spectra of the galaxy corresponding to locations marked as A, B and N in the IFU image and extracted within apertures of 0\farcs2\,$\times\,$0\farcs2. The location of the nucleus has been assumed to be the location of the peak of the continuum emission. We adopt an inclination of the disk of 40\ensuremath{^\circ} from the axial ratio (from NED).

Spectra N and B are typical of Seyfert 2 galaxies, showing [N\,II]\,$\lambda$$\lambda$6548,6583, H$\alpha$ and [S\,II]\,$\lambda$$\lambda$6717,6731 emission lines, with He\,I\,$\lambda$5876, [Fe\,VII]\,$\lambda$6086 and [O\,I]\,$\lambda$$\lambda$6300,6363 also present in the spectrum of the nucleus (N). The spectrum from location A shows narrow (velocity dispersion of $\approx$\,60\,km\,s$^{-1}$) emission lines and an [N\,II]/H$\alpha$ ratio of about 0.3, characteristic of H\,II regions. The spectrum from region N shows asymmetric emission lines with blue wings, visible particularly in [FeVII] and H$\alpha$, where the wing causes the line to blend with [N\,II]\,$\lambda$6548. The spectrum from region B shows lines with a velocity dispersion of $\approx$\,100\,km\,s$^{-1}$ and [N\,II]/H$\alpha$\,$>$\,1, typical of the extended narrow line region of AGNs.

\subsection{Measurements}

\begin{figure*}
\includegraphics[scale=1.19]{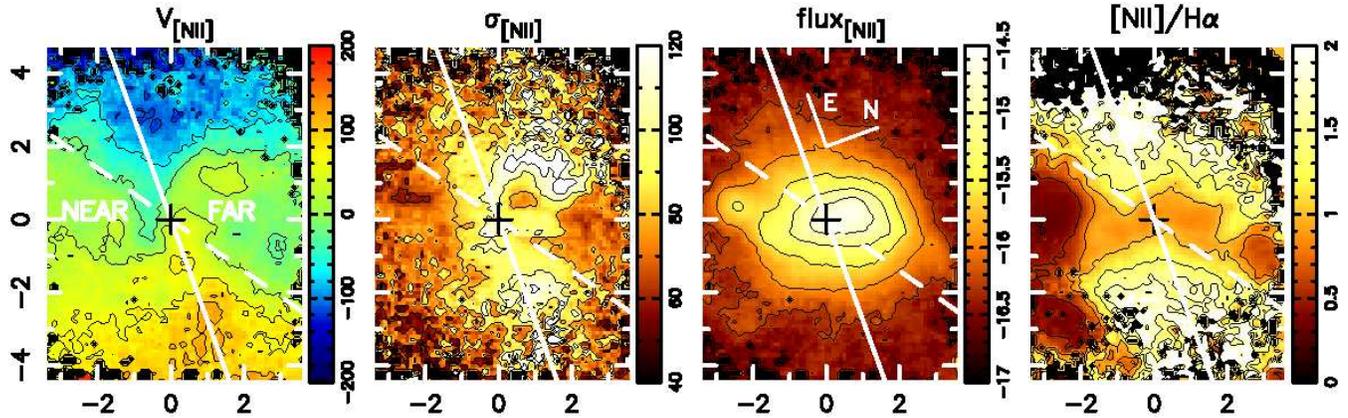}
\caption{Centroid velocity (km\,s$^{-1}$), velocity dispersion (km\,s$^{-1}$), flux distribution in logarithmic scale (erg\,cm$^{-2}$\,s$^{-1}$ per pixel) and [N\,II]/H$\alpha$ maps as obtained by fitting a single Gaussian to the [N\,II] and H$\alpha$ lines. The solid white line marks the position of the line of nodes, the dashed line marks the position of the nuclear bar and the black cross marks the position of the nucleus, assumed to correspond to the peak of the continuum emission.}
\label{fig4}
\end{figure*}

\begin{figure*}
\includegraphics[scale=1.49]{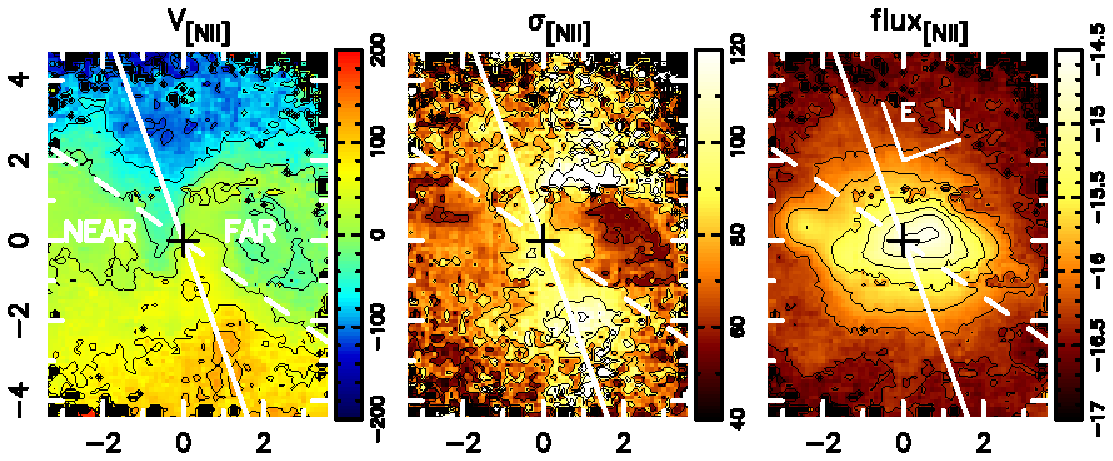}
\includegraphics[scale=1.49]{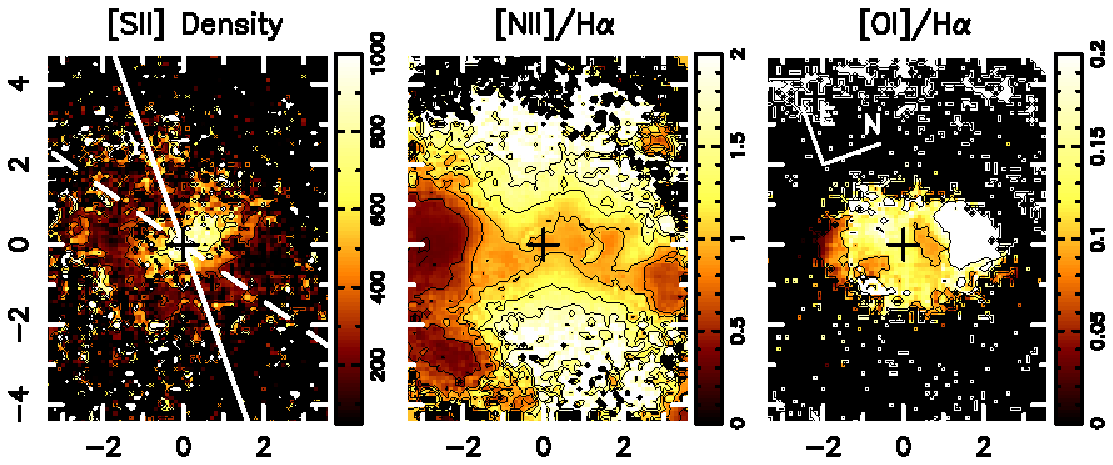}
\caption{Top panel: centroid velocity (km\,s$^{-1}$), velocity dispersion(km\,s$^{-1}$) and flux distribution in logarithmic scale (erg\,cm$^{-2}$\,s$^{-1}$ per pixel) of the narrower component.  Bottom panel: electron density (cm$^{-3}$), [N\,II]/H$\alpha$ and [O\,I]/H$\alpha$ of the narrower component. The solid white line marks the position of the line of nodes, the dashed line marks the position of the nuclear bar and the black cross marks the position of the nucleus, assumed to correspond to the peak of the continuum emission.}
\label{fig5}
\end{figure*}

\begin{figure*}
\includegraphics[scale=1.49]{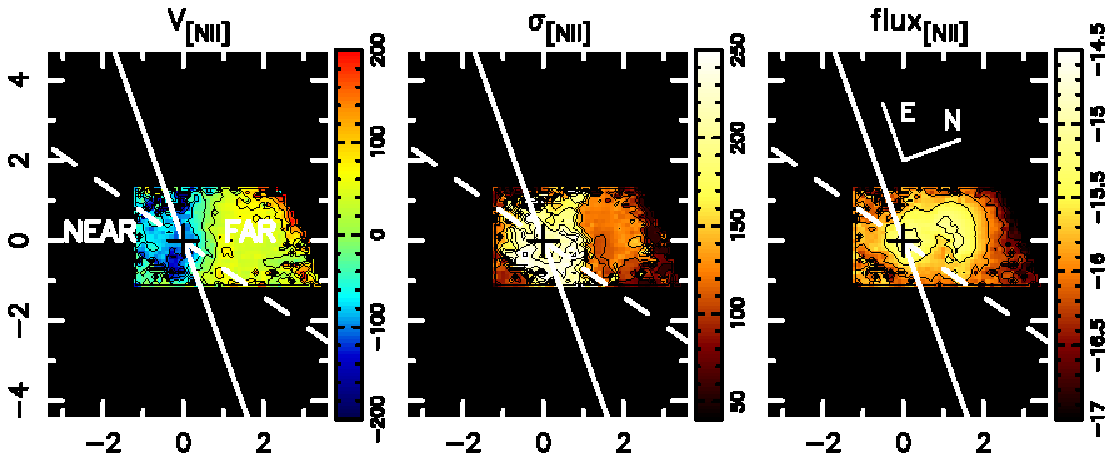}
\includegraphics[scale=1.49]{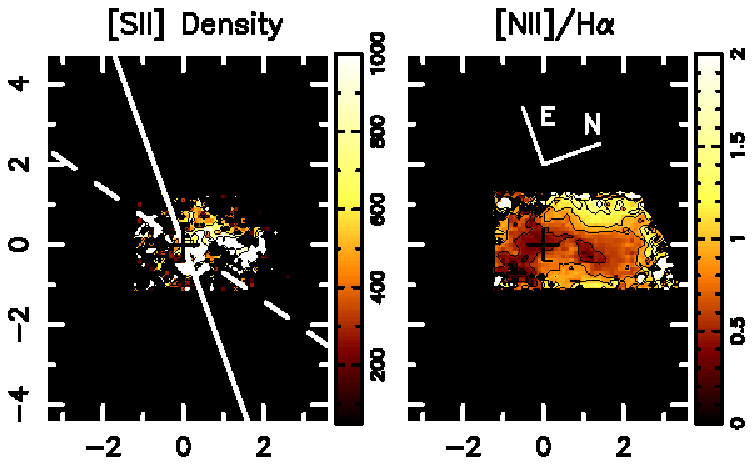}
\caption{Top panel: centroid velocity (km\,s$^{-1}$), velocity dispersion(km\,s$^{-1}$) and flux distribution in logarithmic scale (erg\,cm$^{-2}$\,s$^{-1}$ per pixel) of the broader component.  Bottom panel: density (cm$^{-3}$) and [N\,II]/H$\alpha$ of the broader component. The solid white line marks the position of the line of nodes, the dashed line marks the position of the nuclear bar and the black cross marks the position of the nucleus, assumed to correspond to the peak of the continuum emission.}
\label{fig6}
\end{figure*}

The gaseous centroid velocities, velocity dispersions and the emission-line fluxes were obtained by fitting Gaussians to the [N\,II], H$\alpha$, [O\,I] and [S\,II], He\,I and [Fe\,VII] emission lines. Although at most locations a single Gaussian fits  the emission-line profiles well, close to the nucleus (see the description below), the profiles are best fitted with two Gaussians. Errors were estimated from Monte Carlo simulations in which Gaussian noise is added to the spectra; 100 iterations were performed. In order to limit the number of free parameters in our fit, we adopted the following physically motivated constraints:

\begin{enumerate}
\item flux$_{[N\,II]\,\lambda6583}$/flux$_{[N\,II]\,\lambda6548}=3$;
\item the [N\,II]\,$\lambda$6583, [N\,II]\,$\lambda$6548 and H$\alpha$ lines have the same FWHM;
\item the [N\,II]\,$\lambda$6583, [N\,II]\,$\lambda$6548 and H$\alpha$ lines have the same centroid velocity.

\end{enumerate} 

In order to determine whether a two-Gaussian fit was necessary to adequately fit the emission lines in a given spectrum, we fitted Gauss-Hermite polynomials to the [N\,II] and H$\alpha$ emission lines and built maps of the $h3$ and $h4$ Gauss-Hermite moments, which parametrize the deviations from Gaussianity, $h3$ being related to the skewness of the profiles and $h4$ to the kurtosis. The respective maps are shown in Fig.\,\ref{fig2}. The Gauss-Hermite moment $h4$ is $\approx$\,0 over most of the inner 2\arcsec, only reaching values higher than 0.1 in regions where the signal-to-noise ratio of the [N\,II]\,$\lambda$6583 line is lower than 10. When the signal-to-noise ratios of [N\,II] are  $\approx$\,10 or lower the values of $h3$ and $h4$ are significantly affected by noise (for example wings might be fitted to noise features adjacent to emission lines influencing the $h3$ values). Accordingly, we only consider the regions where the signal-to-noise ratio of the [N\,II]\,$\lambda$6583 is larger than 10 in our analysis. The $h3$ map, on the other hand, shows the presence of both blue wings ($h3$\,$<$\,0) and red wings ($h3$\,$>$\,0) in the [N\,II] and H$\alpha$ profiles in a region extending from $-$1\farcs5 to 3\arcsec\ in the vertical direction and from -2\arcsec\ to 2\arcsec\ in the horizontal direction. We therefore performed two Gaussian fits within this region. Considering that the typical uncertainty in the flux distribution measurements in the inner 2\arcsec\ is of the order of 10$\%$ (see Section\,\ref{uncertainties}), we discarded as poor fits all those where the ratio between the narrow and broad components flux distribution was lower than 0.1. Two Gaussian fits were performed for all the emission lines present in the spectra when the ratio was larger than 0.1. In the case of He\,I\,$\lambda$5876 and [O\,I]\,$\lambda$6300 no location satisfied the above criteria and only one component was fitted. Four examples of the two Gaussian fit to the [N\,II] and H$\alpha$ lines are shown in Fig.\,\ref{fig3}. As seen in the example spectra, the [N\,II] and H$\alpha$ line profiles are fitted by a ``narrow" component (60\,km\,s$^{-1}$\,$\le$\,$\sigma$\,$\le$\,120\,km\,s$^{-1}$), which fits the line core, and a ``broad component" (100\,km\,s$^{-1}$\,$\le$\,$\sigma$\,$\le$\,250\,km\,s$^{-1}$), which fits the wings. We will call these components ``narrower component" and ``broader component" from now on to avoid confusion, as the terms ``narrow component" and ``broad component" traditionally refer to gas emission originating in the narrow and broad line regions.
\begin{figure*}
\includegraphics[scale=1.49]{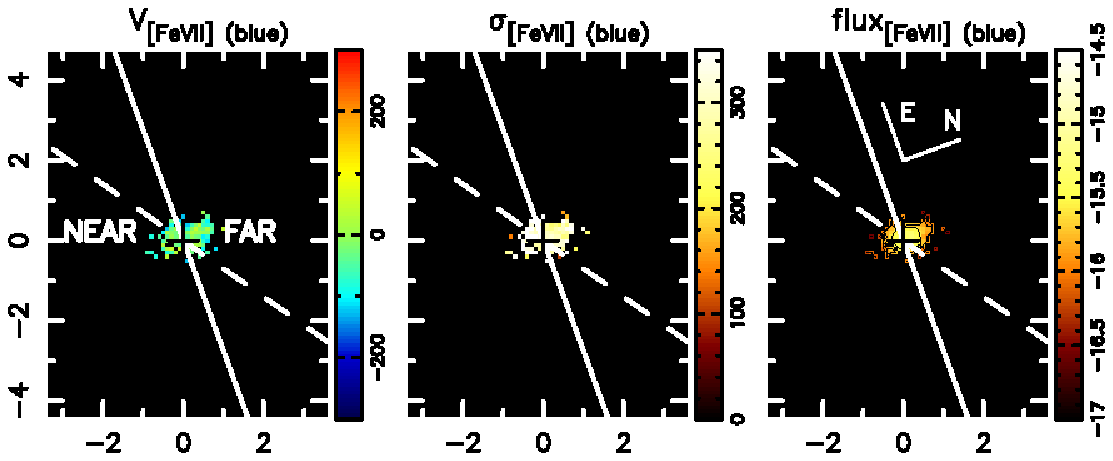}
\includegraphics[scale=1.49]{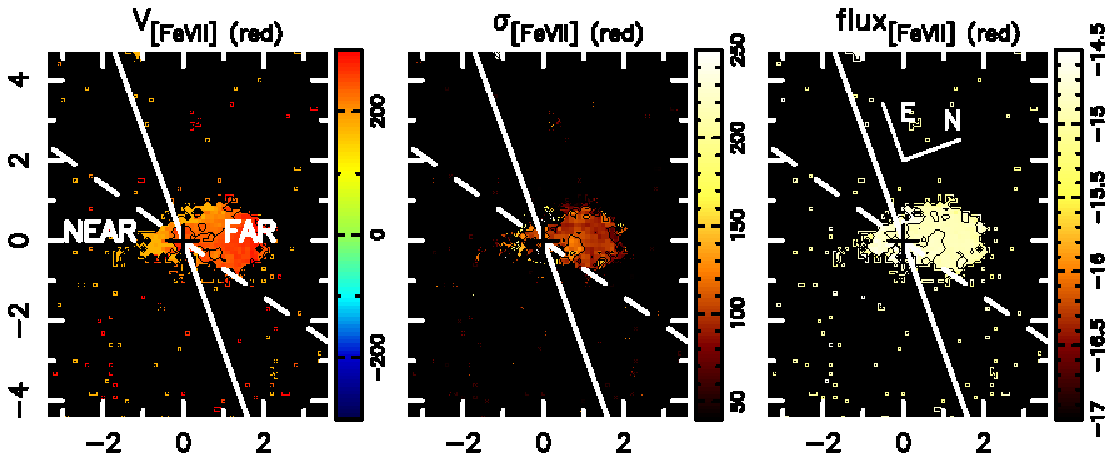}
\caption{Top panel: centroid velocity (km\,s$^{-1}$), velocity dispersion(km\,s$^{-1}$) and flux distribution in logarithmic scale (erg\,cm$^{-2}$\,s$^{-1}$ per pixel) of the [Fe\,VII] blueshifted component. Bottom panel: centroid velocity (km\,s$^{-1}$), velocity dispersion(km\,s$^{-1}$) and flux distribution (10$^{-17}$\,erg\,cm$^{-2}$\,s$^{-1}$ per pixel) of the [Fe\,VII] redshifted component. The solid white line marks the position of the line of nodes, the dashed line marks the position of the nuclear bar and the black cross marks the position of the nucleus, assumed to correspond to the peak of the continuum emission.}
\label{fig7}
\end{figure*}

\begin{figure*}
\includegraphics[scale=1.49]{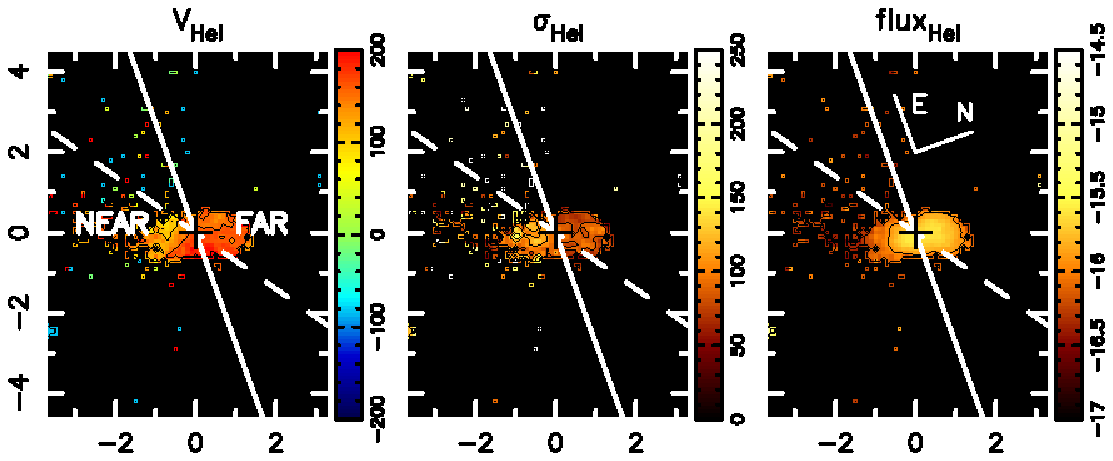}
\caption{Centroid velocity (km\,s$^{-1}$), velocity dispersion(km\,s$^{-1}$) and flux distribution in logarithmic scale (erg\,cm$^{-2}$\,s$^{-1}$ per pixel) of the He\,I emission line. The solid white line marks the position of the line of nodes, the dashed line marks the position of the nuclear bar and the black cross marks the position of the nucleus, assumed to correspond to the peak of the continuum emission.}
\label{fig8}
\end{figure*}

\subsection{Gaseous kinematics}

As the single-Gaussian fit provides a good representation of most of the flux in the profile everywhere (leaving only wings not well fitted in the inner part of the FOV), we present the results from both the one-component and two-component fits.

\subsubsection{Single-component fit}

In Fig.\,\ref{fig4}, we show centroid velocity, velocity dispersion, flux distribution and [N\,II]/H$\alpha$ ratio maps obtained by fitting a single-Gaussian to the [N\,II] and H$\alpha$ lines. A systemic velocity of 2394\,km\,s$^{-1}$ (see Section\,\ref{model}) was subtracted from the centroid velocity maps. The [S\,II] and [O\,I] maps are similar to the [N\,II] maps and thus are not shown. The resulting velocity field shows a distorted rotation pattern in which the east side of the galaxy is approaching and the west side is receding. Under the assumption that the nuclear spiral arms are trailing, it can be concluded that the near side of the galaxy is the south, and the far side is the north. 

The velocity dispersion map shows velocity dispersions of the order of 100\,km\,s$^{-1}$ around the major axis and in a region extending to the northeast. The highest velocity dispersions are observed in a region extending from $\approx$\,1\arcsec\ to $\approx$\,2\arcsec\ north. Over the rest of the FOV, the velocity dispersion is of the order of 80\,km\,s$^{-1}$.

\subsubsection{Two-component fit} 

In Fig.\,\ref{fig5} and Fig.\,\ref{fig6}, we show centroid velocity, velocity dispersion, flux distribution, density and line ratio maps obtained by fitting two Gaussians to the [N\,II], H$\alpha$ and [S\,II] lines. In Fig.\,\ref{fig5}, we show the maps obtained for the narrower component, which were constructed by combining results from both the single and two Gaussian fits. The narrower component centroid velocity and velocity dispersion maps are similar to the single Gaussian maps, although the low velocity dispersion region just to the north of the nucleus is more extended in the narrower component map. The broader component centroid velocity map (Fig.\,\ref{fig6}) shows blueshifted velocities in the inner 1\arcsec. Redshifted velocities are observed to the north of the nucleus, in the far side of the galaxy, reaching 60\,km\,s$^{-1}$. The highest velocity dispersions are observed at the locations showing the highest blueshifts, where values between 160\,km\,s$^{-1}$ and 250\,km\,s$^{-1}$ are observed. To the north of the nucleus the velocity dispersions are lower, between 100\,km\,s$^{-1}$ and 150\,km\,s$^{-1}$.

The [Fe\,VII]\,$\lambda$6086 emission line is clearly double peaked over the inner 0\farcs6 around the nucleus. The blue peak appears as an extended blue wing in spectrum N in Fig\,\ref{fig1}. In Fig.\,\ref{fig7}, we show the maps resulting from fitting two Gaussians to the [Fe\,VII] emission line. The two [Fe\,VII] components are systematically blueshifted and redshifted with respect to the systemic velocity of the galaxy (see Section\,\ref{model}). The [Fe\,VII] blueshifted component shows velocities of about $-$100\,km\,s$^{-1}$ and velocity dispersions of the order of 300\,km\,s$^{-1}$. The [Fe\,VII] redshifted component is observed in a region extending from 1\arcsec\ southeast of the nucleus to $\approx$\,2\arcsec\ northwest of the nucleus. The redshifted component shows velocities within 100--200\,km\,s$^{-1}$ and velocity dispersions within 100-150\,km\,s$^{-1}$. The He\,I centroid velocities and velocity dispersions displayed in Fig.\,\ref{fig8} are similar to those of the [Fe\,VII] redshifted component.

\subsection{Line fluxes and gas excitation}\label{lineflux}

In this section, we discuss only the line fluxes and gas excitation of the two-component fits, as the results from the single-component fit are mostly the same as the ones from the narrower component fit.

The flux distribution of the narrower [N\,II] component is shown in the upper right panel of Fig.\,\ref{fig5}. In the inner 1\arcsec, the isoflux contours are extended approximately along the north-south direction. Between 1\arcsec\ and 2\arcsec\ the contours are extended approximately along the N-S direction and from 2\arcsec\ to the borders of the FOV the contours are oriented along a PA of $\approx$\,160. The H$\alpha$, [S\,II] and [O\,I] flux distribution maps all show similar features; however, the [O\,I] emission is confined within a radius of $\approx$\,2\arcsec.  

The narrower component [N\,II]/H$\alpha$ map has a peculiar shape, with the lowest values observed S, SW and NW of the nucleus. The line ratio values increase with distance from the nucleus along the east-west direction. In the inner 2\arcsec, the [N\,II]/H$\alpha$ line ratio has values between 0.8 and 1.5. There are three regions, one to the northwest, one to the southeast and another to the south of the nucleus, towards the borders of the FOV, that show [N\,II]/H$\alpha$ ratios between 0.2 and 0.5, characteristic of H\,II regions. Similarly, values of the order of 0.6--0.7 are observed in a region 2\arcsec\ northwest of the nucleus, also indicating that this is an H\,II region. The [O\,I]/H$\alpha$ ratio reaches a maximum of 0.2 at $\approx$\,2\arcsec\ north of the nucleus and a minimum of 0.05 at $\approx$\,2\arcsec\ southeast of the nucleus.        

The narrower component gas density map (Fig.\,\ref{fig5}), obtained from the [SII]\,$\lambda\lambda$6717/6731 line ratio assuming an electronic temperature of 10000K \citep{osterbrock06}, has its peak in a region extending to 1\arcsec\ north of the nucleus, where it reaches 800-900\,cm$^{-3}$. The density decreases with radius, reaching a minimum of 200-300\,cm$^{-3}$ at 1\farcs5 from the nucleus. 

Similarly to the narrower component emission, the broader component emission is extended towards the north. The [N\,II]/H$\alpha$ ratio is also similar to the narrower component ratio, with the lowest values occurring S and NW of the nucleus and the ratio increasing along the NE-SW direction. The broader component density varies between 400\,cm$^{-3}$ and 1000\,cm$^{-3}$.   

The [Fe\,VII] red component emission to the southeast of the nucleus is stronger by about a factor of 3 compared to the emission to the northwest. The He\,I flux distribution map is similar to that of the narrower [N\,II] component, showing an extension to the north.

\begin{figure*}
\includegraphics[scale=1.0]{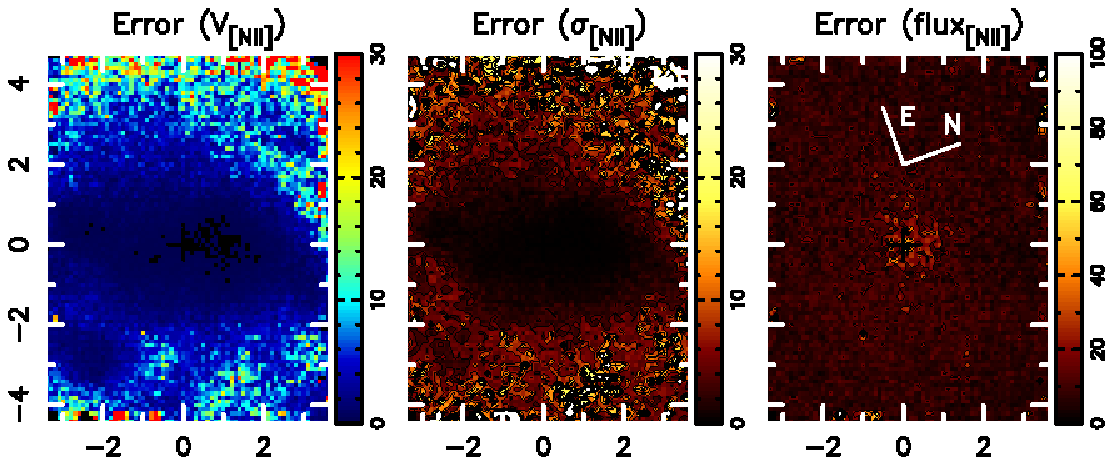}
\includegraphics[scale=1.49]{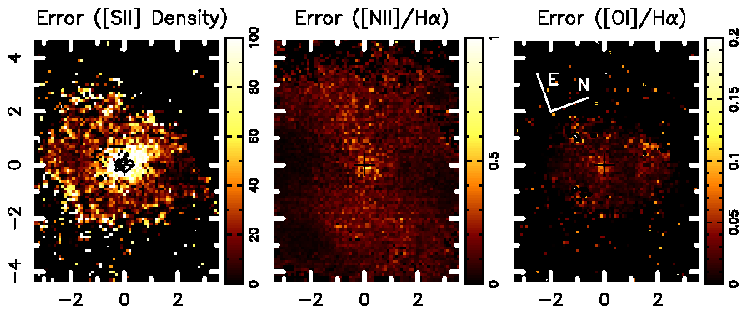}
\caption{Top: uncertainties (\%) in the centroid velocity (km\,s$^{-1}$), velocity dispersion(km\,s$^{-1}$) and flux distribution for the narrower component. Bottom: uncertainties in the elcetron density, [N\,II]/H$\alpha$ and [OI]/H$\alpha$ ratios for the narrower component.}
\label{fig9}
\end{figure*}

\begin{figure*}
\includegraphics[scale=1.0]{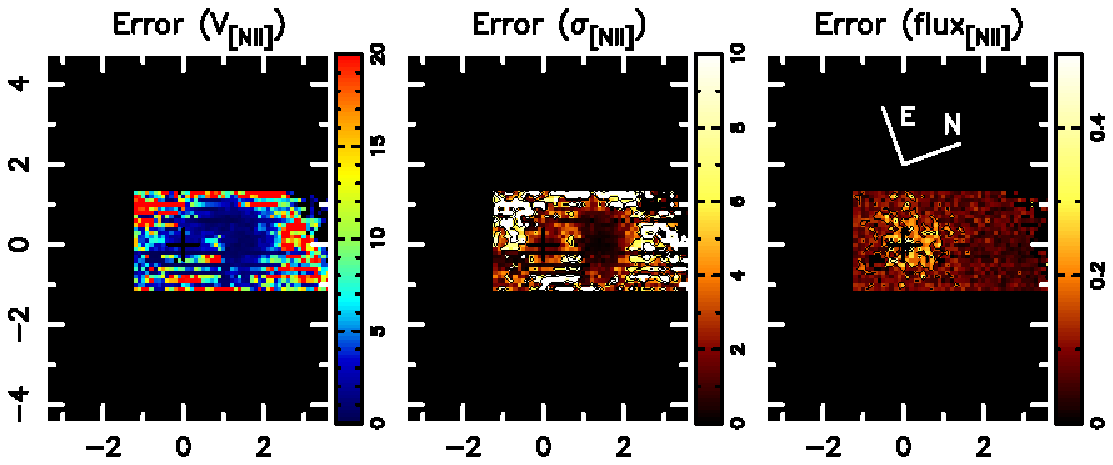}
\includegraphics[scale=1.49]{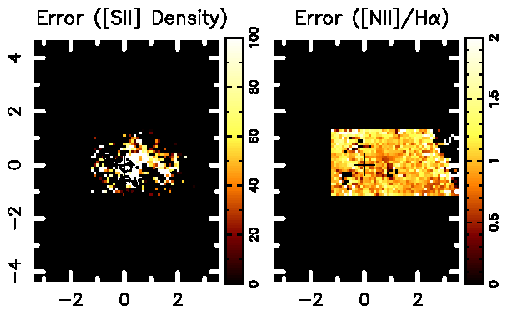}
\caption{Top: uncertainties (\%) in the  centroid velocity (km\,s$^{-1}$), velocity dispersion(km\,s$^{-1}$) and flux distribution for the broader component. Bottom: uncertainties in the electron density and [N\,II]/H$\alpha$ ratio for the broader component.}
\label{fig10}
\end{figure*}

\begin{figure*}
\includegraphics[scale=1.0]{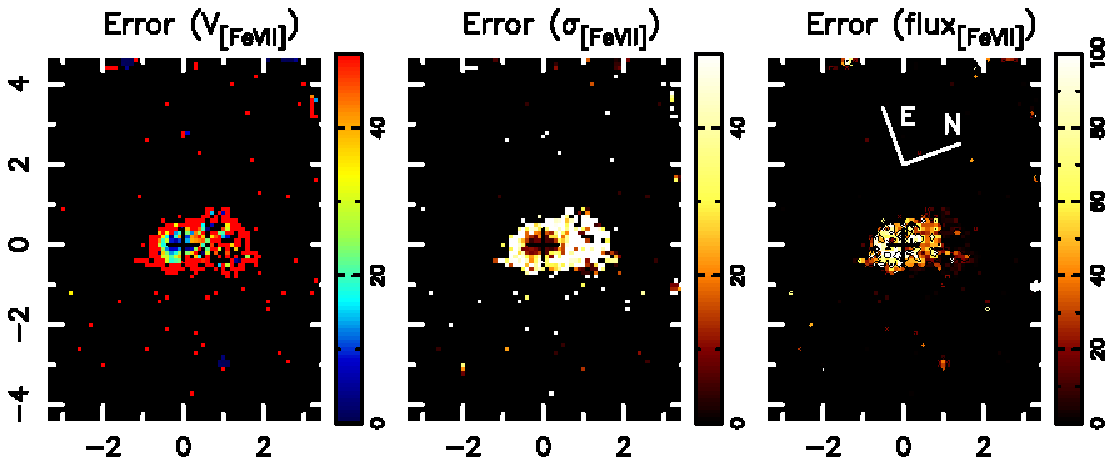}
\includegraphics[scale=1.0]{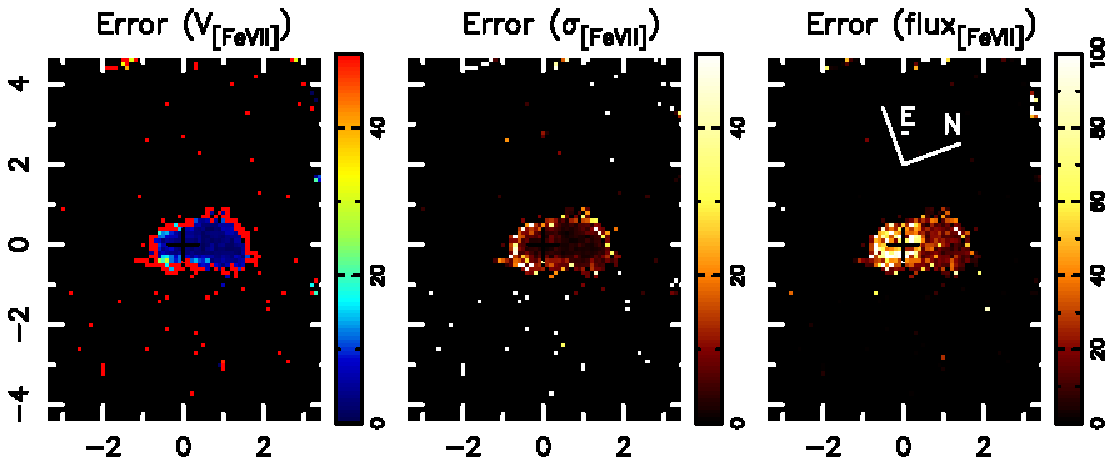}
\includegraphics[scale=1.0]{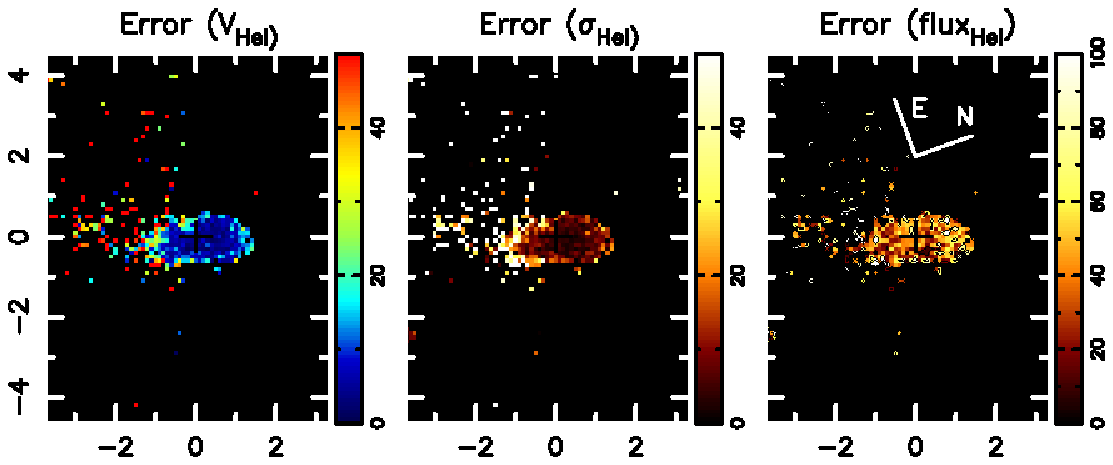}
\caption{Top: uncertainties (\%) in the centroid velocity (km\,s$^{-1}$), velocity dispersion(km\,s$^{-1}$) and flux distribution for the blueshifted [Fe\,VII] component. Middle: uncertainties (\%) in the centroid velocity (km\,s$^{-1}$), velocity dispersion(km\,s$^{-1}$) and flux distributions for the redshifted [Fe\,VII] component. Bottom: Top: uncertainties (\%) in the centroid velocity (km\,s$^{-1}$), velocity dispersion(km\,s$^{-1}$) and flux distributions for the He\,I emission line.}
\label{fig11}
\end{figure*}

\subsection{Uncertainties}\label{uncertainties}

To test the robustness of the fits and estimate the uncertainties in the quantities measured from each spectrum in our datacube, we performed Monte Carlo simulations in which Gaussian noise was added to the observed spectrum. For each spaxel, the noise added in each Monte Carlo iteration was randomly drawn from a Gaussian distribution whose dispersion was set to the expected Poissonian noise of that spaxel. 100 iterations were performed and the estimated uncertainty in each parameter - line center, line width, and total flux in the line - was derived from the $\sigma$ of the parameter distributions yielded by the iterations. In Figs.\,\ref{fig9}, \ref{fig10} and \,\ref{fig11} we show the uncertainties in the measurement of the emission lines.
\section{discussion}\label{Discussion}

\subsection{The broader component}

The velocity map of the broader component (Fig.\,\ref{fig6}) shows blueshifted velocities around the nucleus and to the southeast, mostly in the near side of the galaxy, and redshifted velocities from 0\farcs6 to 3\arcsec\ northwest of the nucleus, in the far side of the galaxy. This kinematics can be interpreted as due to a nuclear outflow extending from 1\farcs2 SSE (215\,pc) to 3\arcsec\ (540\,pc) NNW and from 1\arcsec (179\,pc) WSW to 1\farcs3 (233\,pc) ENE. 

\begin{figure*}
\includegraphics[scale=1.19]{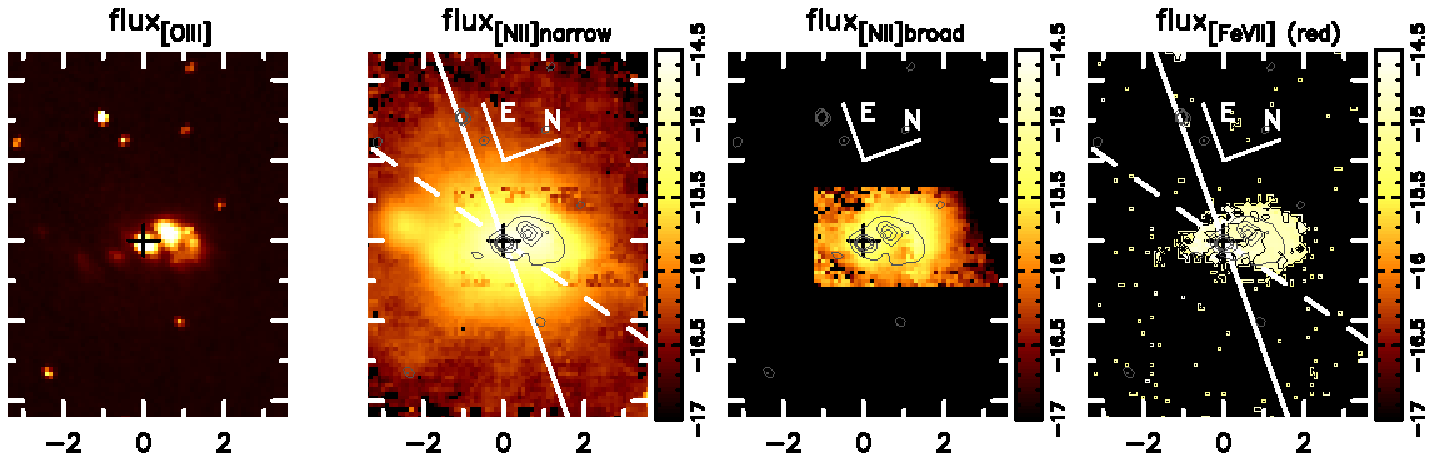}
\caption{Left: [O\,III]\,$\lambda$5007 image from HST observations (WFPC2-PC/F502N, proposal ID: 6419, PI: A. Wilson). Center left: [O\,III] contours overlaid on the [N\,II] narrower component flux distribution. Center right: [O\,III] contours overlaid on the [N\,II]  broader component flux distribution. Right: [O\,III] contours overlaid on the [Fe\,VII]  redshifted component flux distribution. The solid white line marks the position of the line of nodes, the dashed line marks the position of the nuclear bar and the black cross marks the position of the nucleus, assumed to correspond to the peak of the continuum emission.}
\label{fig12}
\end{figure*}

In Fig.\,\ref{fig12}, we present a comparison between the [O\,III]\,$\lambda$5007 flux distribution from HST observations \citep{ferruit00} and both the [N\,II] narrower and broader component flux distributions. The spatial distribution of the [N\,II] broader component emission is similar to that of the [O\,III] emission, so they are likely tracing the same gas. Considering this, we can compare our observations with the HST/STIS (Space Telescope Imaging Spectrograph) spectroscopy slitless observations along PA\,=\,335\ensuremath{^\circ} of the [O\,III]\,$\lambda$5007 line presented by \citet{ruiz05}. They observed blueshifted velocities up to $\approx$\,--150\,km\,s$^{-1}$ at 1\arcsec\ southeast of the nucleus (see Fig.\,4 of their paper) and redshifted velocities up to $\approx$\,150\,km\,s$^{-1}$ at 1\arcsec\ northwest of nucleus. Between 1\arcsec and 2\arcsec northwest of the nucleus, the redshifted velocities decrease to $\approx$\,100\,km\,s$^{-1}$. The blueshifted velocities observed by \citet{ruiz05} are in good agreement with our observations. On the other hand, in a region extending from the nucleus to 1\arcsec northwest, where they observe only redshifted velocities, we observe a mix of blueshifted and redshifted velocities, ranging from $\approx$\,--100\,km\,s$^{-1}$ to $\approx$\,60\,km\,s$^{-1}$. \citet{ruiz05} also observe a mix of blueshifted and redshifted velocities, but farther from the nucleus, between $\approx$\,0\farcs8 and 2\arcsec northwest of the nucleus. Nevertheless, the biggest difference between the GMOS and HST observations is in the nucleus, where \citet{ruiz05} observe velocities around 0\,km\,s$^{-1}$ while we observe blueshifted velocities of $\approx$\,100\,km\,s$^{-1}$. \citet{ruiz05} adopted a systemic velocity of 2459\,km\,s$^{-1}$ in their work, larger than that used here 2394\,km\,s$^{-1}$. Clearly the difference between systemic velocities cannot explain the disagreement between observations, as subtracting a larger velocity would increase the blueshift. Our data are seeing limited, however, and at least part of the differences in the observed velocities could be related to this. It is also worth noting that in the \citet{ruiz05} measurements there are emission structures at the same position with different velocities and velocity dispersions, on both sides of the nucleus, indicating that two velocity components are present. It is likely that these components are the narrower and broader components we identified in our data.      

Finally, considering the similarities between our observations and the [O\,III] observations by \citet{ruiz05}, we conclude that the broader component is due to gas in a bipolar outflow, oriented roughly along the south--north direction. We also argue that the part of the outflow observed in redshift to the north is in front of the galaxy plane, and can thus be observed farther from the nucleus than the blueshifted part of the outflow which comes from behind the galaxy plane and is therefore attenuated. As can be seen in the rightmost panel of Fig.\,\ref{fig12}, the [O\,III] emission traces well the emission of the redshifted [Fe\,VII] component. This indicates that the [Fe\,VII] emission, and presumably the He\,I emission as well, comes from the nuclear outflow.  

The [N\,II]/H$\alpha$ ratio map shows that the lowest values, between 0.2 and 0.8 are observed roughly NE and SW of the nucleus, cospatial with the higher densities. The highest [N\,II]/H$\alpha$ ratios are observed approximately 1\arcsec (179\,pc) east and west of the nucleus and 2\arcsec (358\,pc) north of the nucleus.

The characteristic properties of the broader component gas, in particular its proximity to the nucleus, electron density (600--1000\,cm$^{-3}$), velocity dispersion (120--250\,km\,s$^{-1}$) and its complex kinematics, lead us to identify it as the narrow line region in  NGC\,3081. The fact that there appears be no AGN-photoionized extended narrow line region suggests that the radiation cone (presumably aligned with the bipolar flow) protrudes at a steep angle from the disk.

\subsubsection{Estimating the mass outflow rate}

Having concluded that the kinematics of the broader component are due to an outflow, we now calculate the mass outflow rate, assuming that the outflow occurs in a bicone. The total ejected mass will be twice the mass crossing each end of the bicone, which can be calculated as:
\begin{equation}
\dot{M}_{out}\,=\,\frac{m_{p}\,v\,L_{H\alpha}}{3\,J_{H\alpha}(T)\,N_{e}\,h}
\end{equation}
where $m_{p}$ is the proton mass, $v$ is the outflowing velocity, $L_{H\alpha}$ is the H$\alpha$ luminosity, $J_{H\alpha}(T)$\,=\,3.534$\,\times\,10^{-25}$\,erg\,cm$^{-3}$\,s$^{-1}$ \citep{osterbrock06}, $N_{e}$ is the electron density and $h$ is the height of the bicone (equivalent to the distance from the nucleus).

We will adopt the density, outflowing velocity and $L_{H\alpha}$ values obtained from the redshifted region and a distance to the nucleus of 1\arcsec\ (158\,pc). The total H$\alpha$ flux is 6.7$\,\times\,$10$^{-14}$\,erg\,cm$^{-2}$\,s$^{-1}$. Using the distance of 37.7 Mpc, we obtain a total luminosity for the outflowing gas of $L_{H\alpha}$\,=\,$11.3\,\times\,$10$^{40}$\,erg\,s$^{-1}$. The average electron density in the outflow is 1350\,cm$^{-3}$ (from the [S\,II] ratio). Adopting a distance from the nucleus (height of the cone) of 1\arcsec, we obtain an average projected outflowing velocity of $\approx$\,90\,km\,s$^{-1}$. This velocity, however, needs to be corrected for the inclination of the outflow. As the complicated geometry of the outflow does not allow us to obtain an estimate of the inclination of the central axis of the bicone, we adopt the inclination of the disk (40\ensuremath{^\circ}) as an upper limit for the inclination of the central axis. With this assumption, we obtain a minimum velocity of 140\,km\,s$^{-1}$. Based on the HST [OIII] flux distribution (Fig.\,\ref{fig12}), we assume an opening angle of the bicone of 30\ensuremath{^\circ}. Considering that the redshifted cone is in front of the disk, we obtain a lower limit of 10\ensuremath{^\circ} for the inclination of the central axis, resulting in a maximum velocity of 520\,km\,s$^{-1}$. We thus obtain a lower limit for the outflow mass rate of $\dot{M}_{out}$\,$\approx$\,1.9\,$\times\,10^{-3}$M$_{\odot}$\,yr$^{-1}$ and an upper limit of $\dot{M}_{out}$\,$\approx$\,6.9\,$\times\,10^{-3}$M$_{\odot}$\,yr$^{-1}$.

\subsection{The narrower component}\label{model}

\subsubsection{Kinematics}

\begin{figure*}
\includegraphics[scale=1.19]{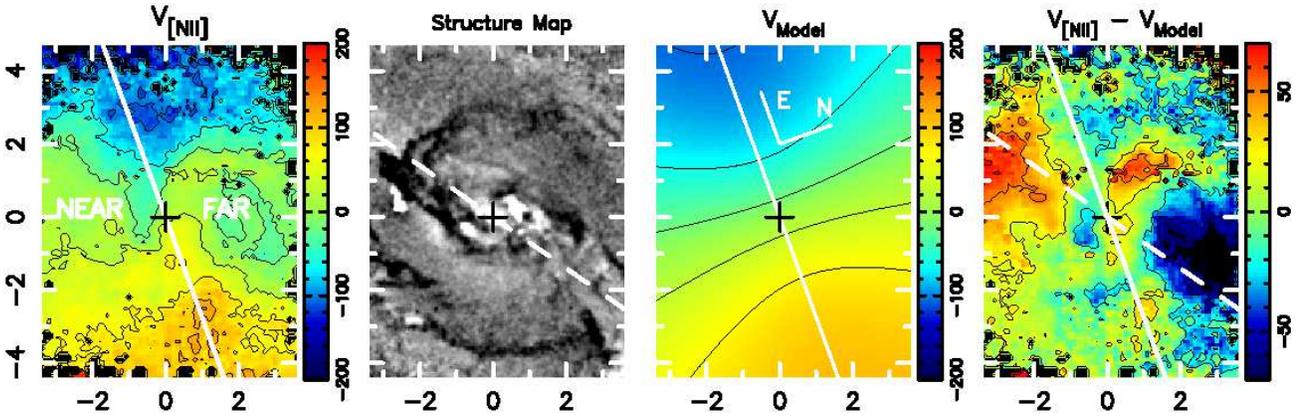}
\caption{Centroid velocity, structure map, modeled velocity field and residuals for the narrower component. The solid white line marks the position of the line of nodes, the dashed line marks the position of the nuclear bar and the black cross marks the position of the nucleus, assumed to correspond to the peak of the continuum emission.}
\label{fig13}
\end{figure*}

As previously mentioned, the velocity map derived from the [N\,II] narrower component covers the whole FOV and shows a typical rotation pattern, although distortions due to non-circular motions are present. In order to isolate these non-circular motions, we modeled the narrower component velocity field assuming a spherical potential with pure circular orbits, with the observed radial velocity at a position ($R,\psi$) in the plane of the sky given by the relation presented in \citet{bertola91}:
\footnotesize
\begin{displaymath}
V=V_{s}+\frac{ARcos(\psi-\psi_{0})sin(\theta)cos^{p}\theta}{\{R^{2}[sin^{2}(\psi-\psi_{0})+cos^{2}\theta cos^{2}(\psi-\psi_{0})]+c^{2}cos^{2}\theta \}^{p/2}} 
\end{displaymath}
\normalsize
where $\theta$ is the inclination of the disk (with $\theta$\,=\,0 for a face-on disk), $\psi_{0}$ is the position angle of the line of nodes, $V_{s}$ is the systemic velocity, $R$ is the radius and $A$, $c,$ and $p$ are parameters of the model. We assumed the kinematical center to be cospatial with the peak of the the continuum emission. We fixed the PA of the kinematic major axis of the gas disk as 90\ensuremath{^\circ} \citep{stoklasova09}, assuming that the gas and stellar disks have the same orientation in the inner 8\arcsec. The inclination was fixed as 40\ensuremath{^\circ}, a value we obtained from the axial ratio (from NED). The resulting parameters $A$, $c$, and $p$ are $250\,\pm$2\,km\,s$^{-1}$, $5\farcs4\pm0.1$ and $1.0\,\pm0.1$ respectively. The systemic velocity corrected to the heliocentric reference frame is $2394\,\pm$6\,km\,s$^{-1}$ (taking into account both errors in the measurement and the fit), in good agreement with previous measurements of the systemic velocity by \citet{theureau05}, who obtained a value of 2391\,km\,s$^{-1}$ from 21cm neutral hydrogen line measurements, and \citet{thaisa96}, who obtained a value of 2385\,km\,s$^{-1}$ from the modeling of the H$\alpha$ velocity curve in the inner 6\arcsec\ along PA\,=\,72.5\ensuremath{^\circ}. The model velocity field is shown in Fig.\,\ref{fig13}, together with the observed velocity field, structure map and residuals.

The residual map (Fig.\,\ref{fig13}) shows that, although there are many regions where the residuals are smaller than 10\,km\,s$^{-1}$, there are two conspicuous regions with high residual velocities along the nuclear bar: a region dominated by redshifts in the near side of the galaxy and a region dominated by blueshifts in the far side, both extending from the borders of the FOV to 1\arcsec\ from the nucleus. Considering the close alignment of these residuals with the axis of the bar, together with the spatial distribution of redshifts and blueshifts, we conclude that they are due to non-circular motions of the gas induced by the gravitational potential of the nuclear bar. The bar can induce shocks, allowing the gas to lose angular momentum and move towards the centre where it can feed the active nucleus of NGC\,3081. This is in line with the ``bar within bars'' scenario proposed by \citet{shlosman89}. This result is similar to that reported by \citet{caceres13}, of possible evidence of gas inflow along nuclear bars in four galaxies, including two Seyfert 2 galaxies and a LINER.

Blueshifted residuals are also observed in a curved strip extending from the borders of the FOV at $\approx$\,3\arcsec north to 2\arcsec\ east, in the far side of the galaxy. A comparison between the residual and structure maps in Fig.\,\ref{fig13} shows that the blueshifted curved strip is cospatial with a dusty nuclear spiral arm, hence we interpret these residuals as due to streaming motions along this nuclear spiral arm. 

A third system of residuals can be identified, consisting of regions of redshifted and blueshifted velocities located 1--2\arcsec NE and $\approx$\,1\arcsec SW of the nucleus, respectively. These residuals could be related to the bipolar outflow observed in the broader component: the redshifted and blueshifted regions closely border the NE and S edges of the broader component flux distribution, respectively (Fig.\,\ref{fig6}). Furthermore, redshifted residuals are associated with a region of increased velocity dispersion 2\arcsec NE of the nucleus. In this scenario, this system results from  an interaction between the bipolar outflow and the disk gas, with the former driving shocks into the disk gas and pushing it outwards. Alternatively, the blueshifted residuals and the redshifted residuals observed between 1\arcsec\ SW and 1\arcsec\ NE can also be due to gas following non-circular orbits in the nuclear bar. This scenario cannot account for the redshifted residuals observed between 1\arcsec\ and 2\arcsec\ NE, however, as they are outside of the bar; thus, in this alternative scenario, part of the redshifted residuals likely originate in an interaction with the nuclear outflow.

\subsubsection{Excitation}

Along the nuclear bar, [N\,II]/H$\alpha$ ratio values between 0.2 and 1.0 are observed. Specifically, cospatial with the redshifted region 2\arcsec\ southeast of the nucleus, the [N\,II]/H$\alpha$ ratio values vary over the range 0.2--0.5, which is typical of H\,II regions. Cospatial to the blueshifted residuals 2\arcsec\ northwest of the nucleus, the [N\,II]/H$\alpha$ ratio values are larger, between 0.6 and 0.7; however, they are still smaller than the values observed closer to the nucleus. We argue that this region is also a site of star formation; the ratio values in this case suggest the presence of a significant fraction of gas that is being photoionized by the AGN and not only by the young stars. This may be evidence that bar-driven gas flows are inducing the formation of new stars in this galaxy. H\,II regions are also observed in the southern nuclear spiral arm, 2\arcsec south of the nucleus. In the inner 2\arcsec\ the [N\,II]/H$\alpha$ ratio values are typical of AGNs. Similar to what is observed in the broader component [N\,II]/H$\alpha$ map, the lowest values, between 0.8 and 1.0, are observed SW and NE of the nucleus. The highest values, between 1.2 and 2, are observed to the east and west of the nucleus, from $\approx$1\arcsec\ to the edges of the FOV. These regions are also where the highest velocity dispersions are observed, which implies that the gas is ionized by shocks. The density map (Fig.\,\ref{fig5}) shows an increase to the northeast, partly cospatial to redshifted residuals NE of the nucleus.

\subsubsection{Estimating an upper limit to the mass inflow rate}

If we assume that the inflow velocity is of the order of the residual velocities observed along the nuclear bar, we can obtain an upper limit to the mass inflow rate. In this case, assuming the total inflow rate is twice that observed in one of the sides of the bar and that the inflow occurs in the plane of the galaxy disk, the ionised gas mass inflow rate which crosses a section of one of the halves of the nuclear bar is given by:
\begin{equation}
\dot{M}_{in}\,=\,N_{e}\,v\,\pi\,r^{2}\,m_{p}\,f
\end{equation}
where $N_{e}$ is the electron density, $v$ is the inflowing velocity of the gas , $m_{p}$ is the mass of the proton, $r$ is the radius of the cross section of the bar and $f$ is the filling factor. The filling factor can be estimated from: 
\begin{equation}
L_{H\alpha}\,\sim\,f\,N_{e}^{2}\,J_{H\alpha}(T)\,V
\end{equation}
where $J_{H\alpha}(T)$\,=\,3.534$\,\times\,10^{-25}$\,erg\,cm$^{-3}$\,s$^{-1}$ \citep{osterbrock06} and $L_{H\alpha}$ is the H$\alpha$ luminosity emitted by a volume $V$. Assuming the volume of the nuclear bar can be approximated by the volume of a cylinder with radius $r$ and height $h$, we obtain: 
\begin{equation}
\dot{M}_{in}\,=\,\frac{m_{p}\,v\,L_{H\alpha}}{J_{H\alpha}(T)\,N_{e}\,h}
\end{equation}

We adopt as inflowing velocity the average residual velocity observed in the redshifted region extending from $\approx$\,1\arcsec\ southeast of the nucleus to the borders of the FOV, 3\farcs3 southeast of the nucleus (see the residual map in Fig.\,\ref{fig13}). After correcting this velocity for the inclination of the galaxy, we obtain an average residual velocity of 59\,km\,s$^{-1}$. As the gas flows along the nuclear bar, we also need to correct this velocity based on the position of the bar in relation to the line of sight. We then obtain a velocity of 111\,km\,s$^{-1}$. The total H$\alpha$ flux and the average density are 5.1$\,\times\,$10$^{-14}$\,erg\,cm$^{-2}$\,s$^{-1}$ and 440\,cm$^{-3}$ respectively. Adopting a distance of 37.7\,Mpc, we obtain $L_{H\alpha}$\,=\,8.6$\,\times\,$10$^{39}$\,erg\,s$^{-1}$. Adopting a height of 2\farcs3, we obtain an upper limit to the mass inflow rate of ionized gas of $\phi$\,$\approx$\,2.4\,$\times\,10^{-2}$M$_{\odot}$\,yr$^{-1}$ at 1\arcsec\ from the nucleus.

The mass accretion rate necessary to produce the luminosity of the Seyfert nucleus of NGC\,3081 is calculated as follows:
\[
\dot{m}\,=\,\frac{L_{bol}}{c^{2}\eta}
\]
with $\eta$\,$\approx$\,$0.1$ (\citealt{frank02} as usually adopted for Seyfert galaxies). Adopting a bolometric luminosity of $L_{bol}$\,=\,4\,$\times\,$10$^{43}$\,erg\,s$^{-1}$ \citep{esquej14} we derive an accretion rate of $\dot{m}$\,=\,7$\,\times\,$10$^{-3}$\,M$_{\odot}$\,yr$^{-1}$. 

The upper limit to the ionized gas mass inflow rate is 10 times larger than the nuclear accretion rate. This is not surprising, as inflow velocities in bars are expected to be much lower than 111\,km\,s$^{-1}$ \citep{athanassoula92,regan04}. An inflow velocity as low as $\approx$\,10\,km\,s$^{-1}$ would drive enough ionized gas inwards to feed the AGN. If there are significant quantities of neutral and molecular gas inflowing, even lower inflow velocities could provide enough gas.

\section{conclusions}\label{Conclusion}

We have measured the gaseous kinematics in the inner 1.2\,$\times$\,1.8\,kpc$^2$ of the Seyfert\,2 galaxy NGC\,3081, from optical spectra obtained with the GMOS integral field spectrograph on the Gemini North telescope at a spatial resolution of $\approx$\,100\,pc. The main results of this paper are as follows:

\begin{itemize}

\item Extended gas emission is observed over the whole FOV, with the profiles being well fitted by a combination of Gaussian curves; 

\item In the inner $\approx$\,2\arcsec, two-components are needed: a narrower component, that is present over the entire FOV, and a broader component, which is present only in the inner 2\arcsec;

\item The broader component centroid velocity map shows blueshifted velocities in the near side of the galaxy and redshifted velocities to the north of the nucleus, in the far side of the galaxy. We interpret this component as a bipolar outflow oriented along the north--south direction. From the measured velocities, fluxes and density of the outflowing gas we estimate a lower limit for the outflow mass rate of $\dot{M}_{out}$\,$\approx$\,1.9\,$\times\,10^{-3}$M$_{\odot}$\,yr$^{-1}$ and an upper limit of $\dot{M}_{out}$\,$\approx$\,6.9\,$\times\,10^{-3}$M$_{\odot}$\,yr$^{-1}$;

\item The narrower component centroid velocity map shows a distorted rotation pattern in which the east side of the galaxy is approaching and the west side is receding. After subtraction of a rotation model, we observe a redshifted region in the near side of the galaxy and a blueshifted region in the far side, both extending from the borders of the FOV to 1\arcsec\ from the nucleus and cospatial with the nuclear bar.

\item  We interpret these residuals as due to non-circular motions induced by the potential of the nuclear bar that may lead to gas inflow. From the measured gas velocities and emission-line fluxes along the bar, we estimate an upper limit to the ionized gas mass inflow rate of $\phi$\,$\approx$\,1.3\,$\times\,10^{-2}$M$_{\odot}$\,yr$^{-1}$.

\item We also observe in the narrower component redshifted residuals extending from the nucleus to the north-northeast, in the far side of the galaxy, and some blueshifted residuals south of the nucleus. These residuals can be due to an interaction between the bipolar outflow and gas in the disk, driving it outwards, or due to non-circular orbits in the barred potential.

\item The low [N\,II]/H$\alpha$ ratios observed along the bar indicate the presence of H\,II there.

\end{itemize}

With our IFU observations, we have disentangled multiple kinematical components contributing to the complex gas velocity field within the inner $\approx$\,1 kpc of this Seyfert 2 galaxy. These include rotation in the galaxy disk plane, a bipolar outflow from the AGN, non-circular motions along the nuclear bar, and an interaction between the bipolar outflow and the disk gas. 

\section*{ACKNOWLEDGMENTS}

We thank the anonymous referee for providing comments and suggestions that have improved this paper. This work is based on observations obtained at the Gemini Observatory, which is operated by the Association of Universities for Research in Astronomy, Inc., under a cooperative agreement with the NSF on behalf of the Gemini partnership: the National Science Foundation (United States), the Science and Technology Facilities Council (United Kingdom), the National Research Council (Canada), CONICYT (Chile), the Australian Research Council (Australia), Minist\'erio da Ci\^encia e Tecnologia (Brazil) and south-eastCYT (Argentina). NN acknowledges funding from ALMA-Conicyt 31110016, BASAL PFB-06/2007, and the FONDAP Center for Astrophysics. This material is based upon work supported in part by the Brazilian institutions CNPq and CAPES and by the National Science Foundation under Award No. AST-1108786. We wish to recognize and acknowledge the cultural role and reverence that the summit of Mauna Kea has always had within the indigenous Hawaiian community. We are most fortunate to have the opportunity to obtain data from observations conducted from this mountain. 

\bibliographystyle{mn2e.bst}
\bibliography{ngc3081.bib}

\label{lastpage}
\end{document}